\documentclass{elsart}

\usepackage{graphicx}

\usepackage{amssymb,amsbsy,amsmath}

\newcommand{\op}[1]{%
    \fontdimen12\textfont3=2pt\fontdimen12\scriptfont3=1.4pt%
    \!\null\mathop{\vphantom{#1}\smash{#1}}\limits_{\sim}\null\!}

\newcommand{\fmref}[1]{(\protect\ref{#1})}
\def\bra#1{\langle \, {#1} \, | \,}
\def\ket#1{\, | \, {#1} \, \rangle}
\newcommand{\braket}[2]{\langle \, {#1} \, | \, {#2} \, \rangle}

\newcommand{\dimhm}[2]{D({#1},{#2})}

\journal{Journal of Computational Physics}

\begin{document}
\begin{frontmatter}

  \title{Optimized implementation of the Lanczos method for magnetic systems}

\author[adr1]{J\"urgen Schnack\corauthref{cor1}}
\address[adr1]{Universit\"at Bielefeld, Fakult\"at f\"ur Physik,
  Postfach 100131, D-33501 Bielefeld, Germany}
\corauth[cor1]{Tel: ++49 521 106-6193; fax: -6455; Email: jschnack@uni-bielefeld.de}

\author[adr2]{Peter Hage \and Heinz-J\"urgen Schmidt}
\address[adr2]{Universit\"at Osnabr\"uck, Fachbereich Physik,
D-49069 Osnabr\"uck, Germany}

\begin{abstract}
  Numerically exact investigations of interacting spin systems
  provide a major tool for an understanding of their magnetic
  properties. For medium size systems the approximate Lanczos
  diagonalization is the most common method. In this article we
  suggest two improvements: efficient basis coding in subspaces
  and simple restructuring for openMP parallelization.
\end{abstract}


\end{frontmatter}

\section{Introduction}
\label{sec-1}

Many magnetic materials can accurately be described by the
Heisenberg or related effective spin models. Due to the vastly
increasing size of the underlying Hilbert space, which grows as
$(2 s + 1)^N$ for $N$ spins of spin quantum number $s$, only
small spin systems can be modeled exactly, i.e. their complete
eigenspectrum can be determined. For larger systems approximate
methods such as the Lanczos \cite{Lan:JRNBS50} or related
methods like the Arnoldi, the projection, or the Density Matrix
Renormalization Group (DMRG) method
\cite{Whi:PRB93,WhD:PRB93B,Sch:RMP05} are used. They usually aim
at properties of ground states in orthogonal subspaces, which
are provided by symmetry, see
e.g. \cite{SHS:PRL02,SNK:PRB04,SSS:PRL05}. But also thermal
properties can be addressed by means of a finite-temperature
Lanczos method \cite{JaP:AP00} as done for instance for the
evaluation of certain Kondo lattice models in
Ref. \cite{ZST:PRB06}.

For all these methods it is of course advantageous to use the
present symmetries in order to reduce the size of the
Hamiltonian matrix as much as possible by decomposing the
Hilbert space into mutually orthogonal subspaces. One obvious
symmetry is the rotational invariance of many models with
respect to rotations about the $z$-axis in spin space. This
leads to a decomposition of the total Hilbert space ${\mathcal
H}$ into orthogonal subspaces ${\mathcal H}(M)$ characterized by
their total magnetic quantum number $M$. The related basis,
which is a subset of the full basis, should then efficiently be
encoded. In nowadays applications these basis states are either
stored in tables and assessed via hash search methods, see
e.g. \cite{PhysRevB.34.1677}, or encoded using the
two-dimensional representation by Lin~\cite{PhysRevB.42.6561},
which needs two vectors of size $\approx (2 s + 1)^{(N/2)}$ for
encoding. In this article we will provide direct algorithms for
encoding and decoding of basis states in subspaces ${\mathcal
H}(M)$.

Thanks to available SMP (symmetric multiprocessing) computers
with large shared memory Lanczos vectors of considerable size
can be processed.  An example is given in Ref.~\cite{SSS:PRL05}
where Lanczos vectors with about $10^9$ entries were used. We
show that by a simple reformulation of the typical
implementation of the Lanczos algorithm a very sufficient
parallelization with openMP can be achieved that avoids write
conflicts.

The article is organized as follows. The next section shortly
introduces the Heisenberg model as an archetypical example. In
section \ref{sec-3} we introduce the new basis encoding in
subspaces ${\mathcal H}(M)$. The last section \ref{sec-4} deals
with parallelization issues.

\section{Heisenberg Hamiltonian and basis encoding}
\label{sec-2}

Spin systems are very often modeled by effective spin
Hamiltonian such as the isotropic Heisenberg Hamiltonian
\begin{eqnarray}
\label{E-2-1}
\op{H}
&=&
-
\sum_{u, v}\;
J_{uv}\,
\op{\vec{s}}(u) \cdot \op{\vec{s}}(v)
\ .
\end{eqnarray}
$\op{\vec{s}}(u)$ are the individual spin operators at sites
$u$. $J_{uv}$ are the matrix elements of the symmetric coupling
matrix. In the following we will assume that all spin quantum
numbers are equal, i.e. $s_1=s_2=\cdots = s_N = s$.

The starting point for any diagonalization is the product basis
of the single-particle eigenstates of all $\op{s}_z(u)$
\begin{eqnarray}
\label{magmol-E-4-5}
\op{s}_z(u)\,
\ket{m_1, \dots, m_u, \dots, m_N}
=
m_u\,
\ket{m_1, \dots, m_u, \dots, m_N}
\ .
\end{eqnarray}
These states are sometimes called Ising states. They span the
full Hilbert space and are used to construct symmetry-related
basis states. For encoding purposes, and since $m_u$ can be
half-integer, they are usually rewritten in terms of quantum
numbers $a_u=s-m_u$ instead of $m_u$, where
$a_u=0,1,\dots,2s$. The number of basis states, i.e. the
dimension of the full Hilbert space, is
$\text{dim}\left({\mathcal H}\right)=(2s+1)^N$. The complete
basis set $\ket{a_1, \dots, a_u, \dots, a_N}$ provides itself a
natural encoding given by the number system with basis $(2s+1)$.
To give an example, the basis of a system of 8 spins $s=1$ can
be completely and easily encoded using all 8-digit numbers where
each digit can assume the values $0,1,2$:
\begin{eqnarray}
\label{E-3-a}
&\ket{0,0,0,0,0,0,0,0}&\\
&\ket{1,0,0,0,0,0,0,0}&\nonumber\\
&\ket{2,0,0,0,0,0,0,0}&\nonumber\\
&\ket{0,1,0,0,0,0,0,0}&\nonumber\\
&\dots&\nonumber\\
&\ket{2,2,2,2,2,2,2,2}&\nonumber
\ .
\end{eqnarray}

\section{Basis encoding in ${\mathcal H}(M)$}
\label{sec-3}

The basis in the subspace ${\mathcal H}(M)$ is given by all
product states $\ket{a_1, \dots, a_N}$ with $M=Ns-\sum_u
a_u$. For usage in a computer program they need to be assigned
to integer numbers $1, \dots, \text{dim}\left({\mathcal
H}(M)\right)$. The reason is that one usually does not need the
basis only once at initialization, but at every Lanczos
iteration, since the sparse Hamiltonian matrix is not stored,
but its non-zero matrix elements are evaluated whenever needed
using
\begin{eqnarray}
\label{E-3-b}
\bra{i}\op{H}\ket{j}
&\equiv&
\bra{a_1^i, \dots, a_N^i}\op{H}\ket{a_1^j, \dots, a_N^j}
\ .
\end{eqnarray}
For a direct coding algorithm of basis states in
subspaces ${\mathcal H}(M)$ it is advantageous that the the
sizes of the subspaces ${\mathcal H}(M)$ are known analytically
\cite{BSS:JMMM00}. Thus an array can be built at startup that
contains for a fixed $s$ the sizes of these subspaces ${\mathcal
H}(M=Ns-A)$ for given $N$ and $A$. We will call this array
$\dimhm{N}{A}$. It will be used to determine the sequential
number of a basis vector in ${\mathcal H}(M)$.  The recursive
buildup is performed using the following relation between the
sizes of subspaces
\begin{eqnarray}
\label{E-3-c}
\dimhm{N}{A}
&=&
\sum_{k=0}^{2s}
\,
\dimhm{N-1}{A-k}
\ ,
\end{eqnarray}
with $\dimhm{N=1}{A=0,1,\dots,2s}=1$, $\dimhm{N}{A=0}=1$, and
$\dimhm{N}{A=1}=N$. If $A \notin \{0,1,\dots,2 N s\}$ then
$\dimhm{N}{A}=0$.

\subsection{$i \Rightarrow \ket{a_1^i, \dots, a_N^i}$}
\label{sec-3-1}

One coding direction, $i \Rightarrow \ket{a_1^i, \dots, a_N^i}$,
which is the more trivial direction, can be realized in several
ways. If the basis is not too big one simply generates all basis
states of the subspace ${\mathcal H}(M)$ in lexicographical
order, compare \fmref{E-3-a}, and stores the quantum numbers
$a_k^i$ of the $i$th vector in an array. The generation can
either be performed by running through all basis states
\fmref{E-3-a} and sorting out those which comply with the
condition $M=Ns-\sum_u a_u$ or by algorithms that generate only
those basis states that obey the condition already.

A direct algorithm $i \Rightarrow \ket{a_1^i, \dots, a_N^i}$
using the known dimensions of the subspaces ${\mathcal
H}(M=Ns-A)$ could be realized as follows\footnote{The given code
uses FORTRAN notation. Nevertheless, it can be easily
transformed into C. One should only pay attention to the fact
that field indices in FORTRAN start at 1 not at 0. Therefore,
the definition of the second field index of $D$ has been
modified accordingly.}
\begin{verbatim}
      m=0
      Ak = A
      do k=N,2,-1
         do n=0,2*s
	    if(i.le.(m+D(k-1,Ak-n+1))) then
	       BasisVector(k) = n
	       Ak = Ak - n
	       goto 100
            else
	       m = m + D(k-1,Ak-n+1)
            endif
         enddo
100      continue
      enddo
      BasisVector(1) = Ak
\end{verbatim}
\verb§BasisVector§ contains the $N$ entries $a_k$. This
algorithm will be made clearer when we explain the inverse
algorithm in subsection \ref{sec-3-2}.

Nevertheless, since a Lanczos routine would run through a state
vector along the lexicographical order of basis states one would
actually only need a function that generates for a given basis
state the succeeding basis state. To understand how this works
it is helpful to picture the basis states $\ket{a_1^i, \dots,
a_N^i}$ as distributions of exactly $A=\sum_u a_u$ balls in $N$
boxes, where each box can contain at most $2s$ balls. Thus the
lexicographically lowest state is given by the distribution
where the boxes are filled sequentially starting with the
leftmost box, i.e. entry number 1.

How does one advance from one basis state to the succeeding one?
\begin{enumerate}
\item Find the leftmost position $k$ for which the entry is
  nonzero and the next entry is less than $2s$. If such a
  position does not exist, then there is no succeeding basis
  state.
\item Take one (ball) out of entry (box) $k$ and add it to the
  next entry to the right, i.e. entry (box) with index $k+1$.
\item Empty all entries (boxes) $1$ to $k$ and fill this content
  (these balls) into the entries (boxes) starting from the left
  in lexicographical order.
\end{enumerate}
Take as an example for $N=8$, $s=3/2$, and $A=6$ the state
$\ket{0,0,0,2,3,1,0,0}$. Entry number $k=5$ from the left is the
first position to fulfill the first condition. One out of the 3
is put into $k=6$ yielding 2 there. Then the content of entries
$k=1,\dots, 5$ is taken and filled into the entries starting
from the left. This content is 4 in the present example. Three
out of the four can be filled into entry number 1. The rest fits
into entry number 2. Therefore, the resulting basis state is
$\ket{3,1,0,0,0,2,0,0}$.

\subsection{$\ket{a_1^i, \dots, a_N^i} \Rightarrow i$}
\label{sec-3-2}

The inverse direction is actually the nontrivial one, since the
basis vectors are only a subset of the full basis set
\fmref{E-3-a}. Therefore, for the latter coding direction search
algorithms are employed, e.g. \cite{PhysRevB.34.1677}, or the
two-dimensional representation of Lin~\cite{PhysRevB.42.6561} is
used, which needs two vectors of size $\approx (2 s +
1)^{(N/2)}$ to encode all basis states.

The position of a basis vector $\ket{a_1, \dots, a_N}$ in the
lexicographically ordered list of vectors will be determined by
evaluating how many vectors lay before this vector. For this
purpose the known dimensions of the subspaces ${\mathcal
H}(M=Ns-A)$ are used again.  We explain this procedure with an
instructive example.  Assume we investigate a spin system with
$N=4$ and $s=3/2$ in a subspace of $A=6$, i.e. $M=0$. Our
example basis vector is $\ket{1,0,2,3}$. In the list of basis
vectors all vectors fulfilling one of the following criteria are
listed before the example vector, the respective dimensions will
be added:
\begin{itemize}
\item Vectors with 0, 1, or 2 instead of 3 as the first
  (rightmost) figure: Their dimensions are $\dimhm{3}{6}$,
  $\dimhm{3}{5}$, and $\dimhm{3}{4}$, respectively, since the
  condition that $\sum_i a_i=A$ must be fulfilled in total.
\item Out of all vectors where the first figure is 3, those
  where the second figure is 0 or 1 are listed before, thus
  their respective dimensions of $\dimhm{2}{3}$ and
  $\dimhm{2}{2}$ must be added.
\item This procedure continues until the last figure. In the
  present example this yields 0 for the third figure and simply
  0 for the last figure.
\item Thus the number of the present vector is given by the sum
  of the mentioned dimensions plus one.
\end{itemize}
In a computer program one can evaluate the position $i$ of
$\ket{a_1, \dots, a_N}$ in the list of basis vectors according
to
\begin{verbatim}
      Ak = A
      i = 1
      do k=N,2,-1
         do n=0,BasisVector(k)-1
            i = i + D(k-1,Ak-n+1)
         enddo
         Ak = Ak - BasisVector(k)
      enddo
\end{verbatim}
\verb§BasisVector§ contains the $N$ entries $a_k$.  If the
array of dimension $\dimhm{N}{A}$ is properly initialized,
i.e. the field value is zero for non-valid combinations of $N$
and $A$, then the sum can be performed in a computer program
without paying attention to the restrictions for the indices.

\section{Parallel Lanczos implementation on SMP machines}
\label{sec-4}

Parallelization of the Lanczos or similar methods aims at a
parallelization of the basic matrix-vector operations. This has
been reported as being extremely difficult due to prohibitive
communication costs \cite{GeR:PC01,ChP:JCP06}. In this section
we show that parallelization is possible if (1) the sparse
matrix is not stored but matrix elements are evaluated whenever
needed and (2) the loops for matrix-vector multiplication are
rearranged.

The basic step of a Lanczos or a similar method consists in the
(repeated) application of the sparse matrix, i.e. the
Hamiltonian, onto an initial trial vector
\begin{eqnarray}
\label{E-4-a}
\braket{i}{\psi_2}
&=&
\sum_{j=1}^{\text{dim}\left({\mathcal H}(M)\right)}\,
\bra{i}\op{H}\ket{j}
\braket{j}{\psi_1}
\ .
\end{eqnarray}
Here $\braket{j}{\psi_1}$ are the entries of the initial column
vector $\ket{\psi_1}$; the resulting vector is $\ket{\psi_2}$.

(1) Although the Hamiltonian matrix $\bra{i}\op{H}\ket{j}$ is
    sparse, it typically contains an order of $N^{1\dots
    2}\times \text{dim}\left({\mathcal H}(M)\right)$ non-zero
    entries, for instance for Heisenberg systems. For very large
    dimensions, e.g. of order $10^9$, this would easily amount
    to several dozens of Gigabytes. Therefore, it would be
    better not to store the matrix, but to evaluate the matrix
    elements whenever needed.

(2) A typical implementation would have the loop about $j$ as
    the outermost loop. An entry $\braket{j}{\psi_1}$ of the
    initial vector would be read, then the non-zero matrix
    elements $\bra{i}\op{H}\ket{j}$ would be determined, and the
    resulting products would be written into the respective
    entries $\braket{i}{\psi_2}$. When parallelizing the loop
    about $j$ this leads to write conflicts since different
    initial entries may result in the same final one. 

It turns out that both problems can be solved together in cases
where the application of the Hamiltonian onto each basis state
is known analytically. In these cases only the non-vanishing
matrix elements will be generated by applying the Hamiltonian,
e.g.  \fmref{E-2-1}, onto the \emph{final} basis state
$\ket{i}$. This yields for a given final index $i$ a set of
initial indices $\{j(i)\}$ where only these indices contribute
in the sum in Eq.~\fmref{E-4-a}.
\begin{eqnarray}
\label{E-4-b}
\braket{i}{\psi_2}
&=&
\sum_{\{j(i)\}}\,
\bra{i}\op{H}\ket{j}
\braket{j}{\psi_1}
\ .
\end{eqnarray}
Therefore, one would rewrite Eq.~\fmref{E-4-a} as
Eq.~\fmref{E-4-b} and in a parallel computer program let $i$ be
the outer loop. Then one determines for every final entry
$\braket{i}{\psi_2}$ those initial entries $\braket{j}{\psi_1}$
that contribute with non-zero $\bra{i}\op{H}\ket{j}$ in the sum
\fmref{E-4-b}. It may happen that at runtime different threads
read the same entry of the initial vector, but this is harmless.

\begin{figure}[!ht]
\begin{center}
\includegraphics[clip,width=65mm]{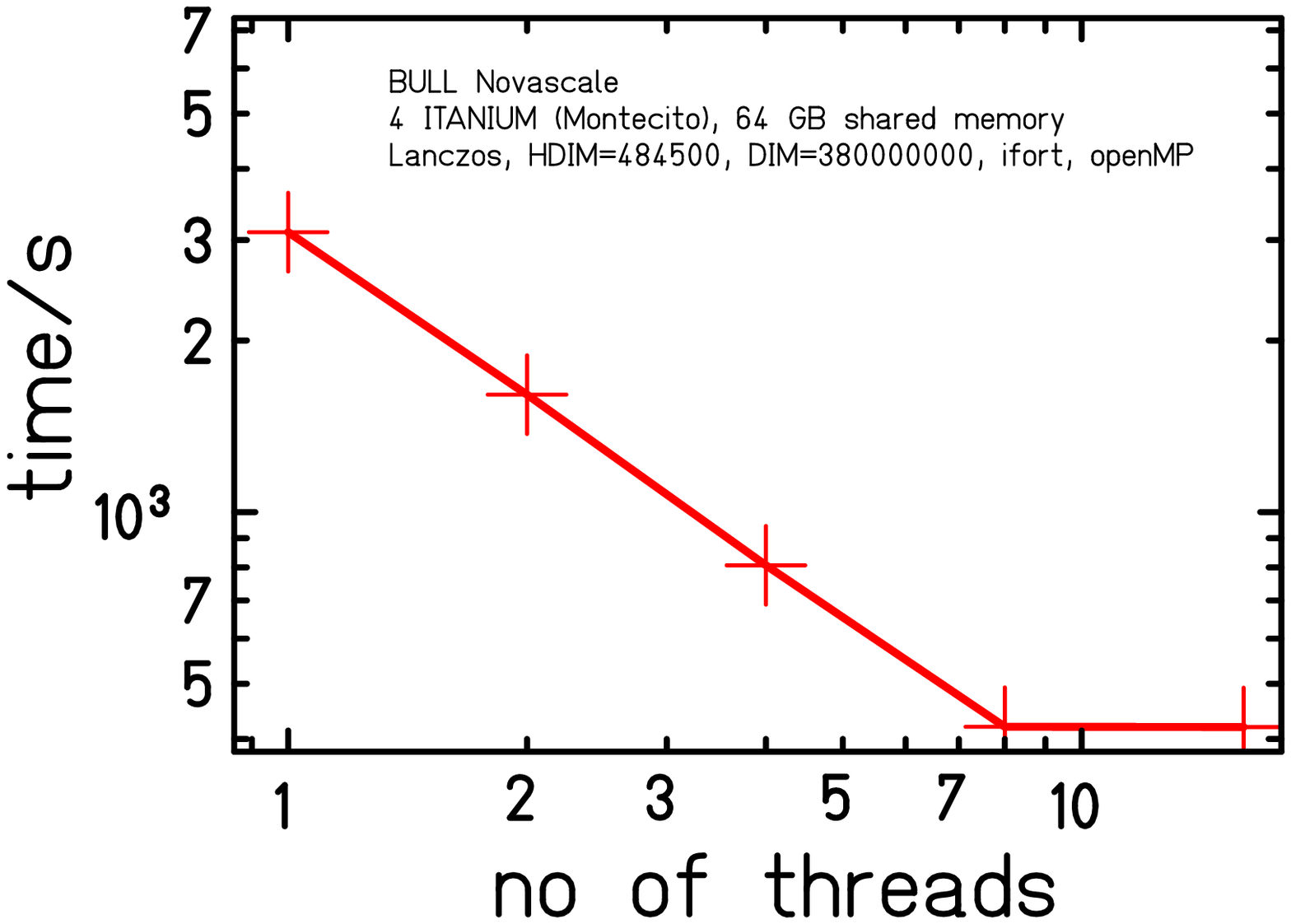}
\quad
\includegraphics[clip,width=65mm]{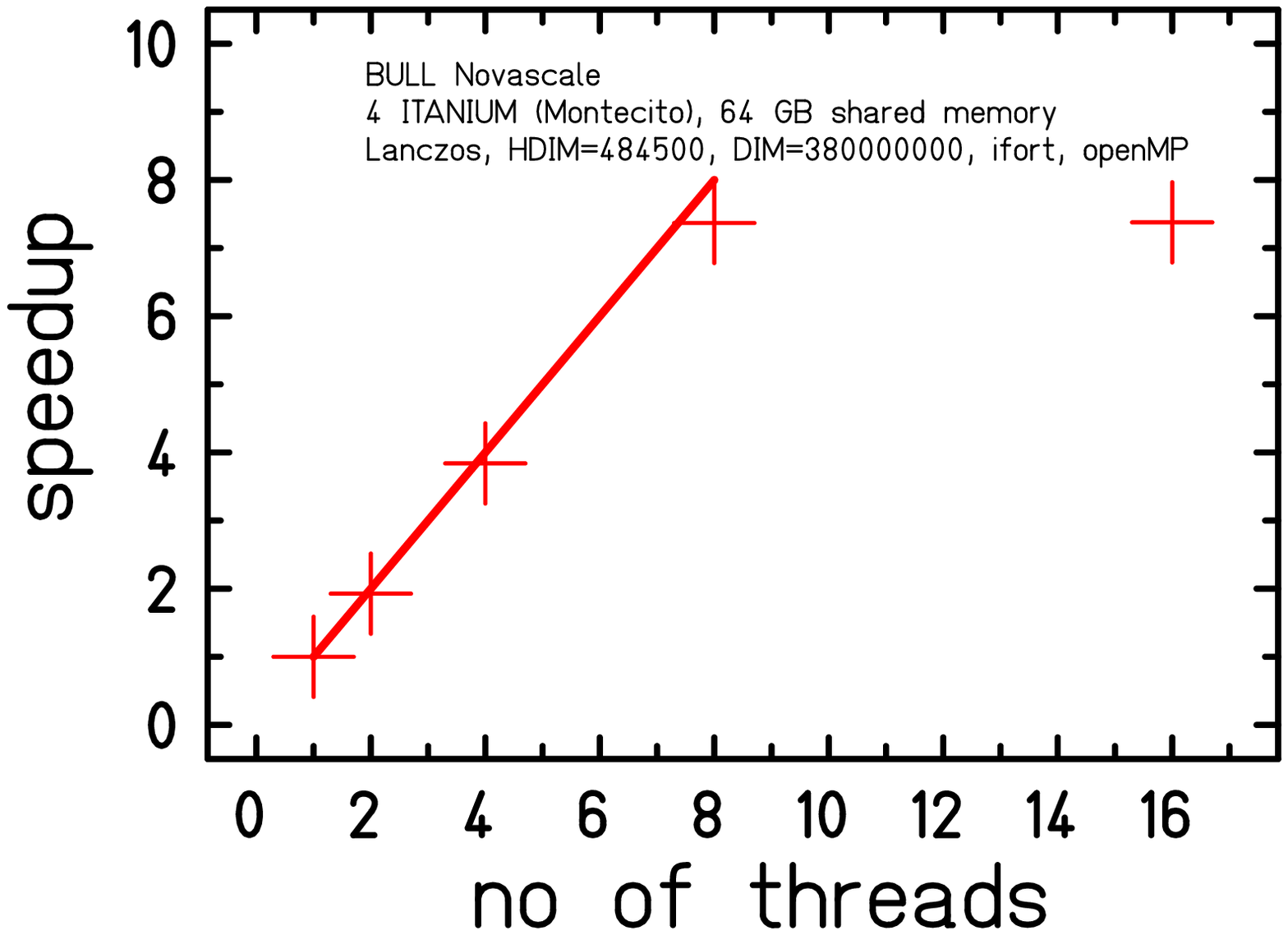}
\vspace*{1mm}
\caption[]{Scaling of CPU time for 200 Lanczos iterations with
  number of allowed threads. The machine has eight cores.}
\label{F-A}
\end{center}
\end{figure}

Figure \ref{F-A} shows as an example the scaling of CPU time for
200 Lanczos iterations with a vector of length $484,500$.  The
program and all subroutines are written in Fortran and compiled
with the INTEL Fortran compiler using openMP directives. The
linear scaling is almost perfect. Slight deviations are due the
non-parallel parts of the program, especially the
initialization.

Summarizing, in this article we provide a coding algorithm for
spin basis states in subspaces ${\mathcal H}(M)$ and
demonstrate that a rearrangement of loops allows an efficient
parallelization of the Lanczos algorithm. The proposed
improvements can easily be ported to similar methods such as
Arnoldi or projection method.

\section*{Acknowledgment}

We thank J.~Schulenburg and B.~Schmidt for discussing their
Lanczos implementations with us. We also thank J.~Richter for
drawing our attention to the encoding of H.Q.~Lin. J.~S. thanks
M.~Br{\"u}ger and R.~Schnalle for discussing encoding problems
with him on a train ride.



\begin{thebibliography}{10}
\expandafter\ifx\csname url\endcsname\relax
  \def\url#1{\texttt{#1}}\fi
\expandafter\ifx\csname urlprefix\endcsname\relax\def\urlprefix{URL }\fi

\bibitem{Lan:JRNBS50}
C.~Lanczos, An iteration method for the solution of the eigenvalue problem of
  linear differential and integral operators, J. Res. Nat. Bur. Stand. 45
  (1950) 255--282.

\bibitem{Whi:PRB93}
S.~R. White, Density-matrix algorithms for quantum renormalization groups,
  Phys. Rev. B 48 (1993) 10345.

\bibitem{WhD:PRB93B}
S.~R. White, D.~Huse, Numerical renormalization-group study of low-lying
  eigenstates of the antiferromagnetic $s=1$ heisenberg chain, Phys. Rev. B 48
  (1993) 3844.

\bibitem{Sch:RMP05}
U.~Schollw\"ock, The density-matrix renormalization group, Rev. Mod. Phys. 77
  (2005) 259--315.

\bibitem{SHS:PRL02}
J.~Schulenburg, A.~Honecker, J.~Schnack, J.~Richter, H.-J. Schmidt, Macroscopic
  magnetization jumps due to independent magnons in frustrated quantum spin
  lattices, Phys. Rev. Lett. 88 (2002) 167207.

\bibitem{SNK:PRB04}
J.~Schnack, H.~Nojiri, P.~K\"ogerler, G.~J.~T. Cooper, L.~Cronin, Magnetic
  characterization of the frustrated three-leg ladder compound
  [(cucl$_2$tachh)$_3$cl]cl$_2$, Phys. Rev. B 70 (2004) 174420.

\bibitem{SSS:PRL05}
C.~Schr{\"o}der, H.-J. Schmidt, J.~Schnack, M.~Luban, Metamagnetic phase
  transition of the antiferromagnetic heisenberg icosahedron, Phys. Rev. Lett.
  94 (2005) 207203.

\bibitem{JaP:AP00}
J.~Jakli{\v c}, P.~Prelov{\v s}ek, Finite-temperature properties of doped
  antiferromagnets, Advances in Physics 49 (2000) 1--92.

\bibitem{ZST:PRB06}
I.~Zerec, B.~Schmidt, P.~Thalmeier, Kondo lattice model studied with the finite
  temperature lanczos method, Phys. Rev. B 73 (2006) 245108.

\bibitem{PhysRevB.34.1677}
E.~R. Gagliano, E.~Dagotto, A.~Moreo, F.~C. Alcaraz, Correlation functions of
  the antiferromagnetic heisenberg model using a modified lanczos method, Phys.
  Rev. B 34~(3) (1986) 1677--1682.

\bibitem{PhysRevB.42.6561}
H.~Q. Lin, Exact diagonalization of quantum-spin models, Phys. Rev. B 42 (1990)
  6561--6567.

\bibitem{BSS:JMMM00}
K.~B{\"a}rwinkel, H.-J. Schmidt, J.~Schnack, Structure and relevant dimension
  of the {H}eisenberg model and applications to spin rings, J. Magn. Magn.
  Mater. 212 (2000) 240.

\bibitem{GeR:PC01}
R.~Geus, S.~Rollin, Towards a fast parallel sparse symmetric matrix-vector
  multiplication, Parallel Comput. 27 (2001) 883--896.

\bibitem{ChP:JCP06}
W.~W. Chen, B.~Poirier, Parallel implementation of efficient preconditioned
  linear solver for grid-based applications in chemical physics. ii: {QMR}
  linear solver, J. Comput. Phys. 219 (2006) 198--209.

\end{thebibliography}

\end{document}